# Large spin to charge conversion in topological superconductor $\beta$-PdBi$_2$ at room temperature

Yang Li[1,3,a)], Shi-jia Yang.[2,a)], Dali Sun[2,6*], Yun-bin Sun[4], Yan Li[1,3], Eric Vetter[2], Rui Sun[1,3], Na Li[1,3], Xu Yang[1,3], Lei Su[1,3], Zi-zhao Gong[1,3], Zong-kai Xie[1,3], Jian-jun Zhao[4], Wei He[1], Xiang-qun Zhang[1,3], and Zhao-hua Cheng[1,3,5,*]

[1]State Key Laboratory of Magnetism and Beijing National Laboratory for Condensed Matter Physics, Institute of Physics, Chinese Academy of Sciences, Beijing 100190, China

[2]Department of Physics, North Carolina State University, North Carolina 27695, USA

[3]School of Physical Sciences, University of Chinese Academy of Sciences, Beijing 100049, China

[4]Key Laboratory of Magnetism and Magnetic Materials at Universities of Inner Mongolia Autonomous Region, Department of Physics, Baotou Normal University, Baotou 014030, China

[5]Songshan Lake Materials Laboratory, Dongguan, Guangdong 523808, China

[6]Organic and Carbon Electronics Laboratory (ORaCEL), North Carolina State University, North Carolina 27695, USA

## ABSTRACT

$\beta$-PdBi$_2$ has attracted much attention for its prospective ability to possess simultaneously topological surface and superconducting states due to its unprecedented spin-orbit interaction (SOC). Whereas most works have focused solely on investigating its topological surface states, the coupling between spin and charge degrees of freedom in this class of quantum material remains unexplored. Here we first report a study of spin-to-charge conversion in a $\beta$-PdBi$_2$ ultrathin film grown by molecular beam epitaxy, utilizing a spin pumping technique to perform inverse spin Hall effect measurements. We find that the room temperature spin Hall angle of Fe/$\beta$-PdBi$_2$, $\theta_{SH} = 0.037$. This value is one order of magnitude larger than that of reported conventional superconductors, and is comparable to that of the best SOC metals and topological insulators. Our results provide an avenue for developing superconductor-based spintronic applications.

a) These authors contributed equally to this work.

*Corresponding authors: zhcheng@iphy.ac.cn or dsun4@ncsu.edu

## I. INTRODUCTION

Spintronics relies on utilizing the electron's spin degree of freedom for the development of future information technologies [1,2]. For these applications, the conversion of spin currents into charge currents and vice versa are key functionalities whose efficiencies are critically determined by the spin-orbit coupling (SOC) in the device materials. Strong SOC stemming from the use of heavy elements, such as commonly used heavy metals like Pt, Ta, or W, allows for the efficient interconversion of charge and spin currents via the spin Hall effect (SHE) and inverse spin Hall effect (ISHE) [3-5]. In these effects, the interconversion is typically accomplished between a three-dimensional (3D) charge current ($J_C^{3D}$) and a 3D spin current ($J_S^{3D}$), in which the interconversion efficiency is given by the spin Hall angle ($\theta_{SH} = J_C^{3D}/J_S^{3D}$), which, in general, is proportional to the SOC strength. Furthermore, higher conversion rates can be achieved by exploiting the SOC-induced properties at the so-called Rashba interfaces and the non-trivial surfaces in topological insulators (TIs) [6-8]. As opposed to the 3D charge current-induced via ISHE, at these momentum-locked topological surfaces or Rashba interfaces a 2D charge current ($J_C^{2D}$) is generated instead via the inverse Rashba-Edelstein effect (IREE), where the conversion efficiency is given by the IREE length ($\lambda_{IREE} = J_C^{2D}/J_S^{3D}$).

Similar to TIs, topological superconductors (TSCs) are a new state of quantum matter with a full pairing gap in the bulk, and topologically protected gapless boundary states that can support exotic Majorana fermions [9,10]. These unique properties make this class of quantum materials attractive for applications ranging from spintronics to quantum computation. Recently, *β*-PdBi$_2$, a special candidate for connate TSC due to its large SOC, has attracted great attention [11-17]. Angle-resolved photoemission spectroscopy (ARPES) revealed the presence of both topological and trivial surface states crossing the Fermi level in the normal state, with in-plane spin polarizations [12-14]. Compared with conventional spin-singlet s-wave superconductors, the superconductivity of *β*-PdBi$_2$ was found to originate from an unconventional pairing symmetry consistent with spin-triplet pairing [15]. Besides the bulk superconducting gap, the highly enhanced full-gap superconductivity in spin-polarized surface states was also observed in the *β*-PdBi$_2$ films [16,17]. Although intensive studies of band structure and superconducting properties in *β*-PdBi$_2$ have been carried out, little attention has been paid to the coupled spin and charge degrees of freedom in heterostructures composed of topological superconductor and magnetic materials, which is of fundamental interest for future applications in superconducting spintronics. Given the intrinsic strong SOC and non-trivial surface states, TSC *β*-PdBi$_2$ is expected to show a high spin to charge conversion efficiency. Accordingly, spin injection into superconductors, typically interpreted in terms of quasiparticle or spin-triplet pair mediated transport, is under active investigation [18-22]. The room temperature spin transport parameters of several conventional superconductors (like Nb and NbN) have also been probed by spin pumping measurements [23,24]. As a result, determining the normal-state spin transport properties in $\beta$-PdBi$_2$ can also be useful to future spintronics research using TSC.

In this work we investigate ferromagnetic resonance (FMR) driven spin pumping in Fe/$\beta$-PdBi$_2$ bilayers to probe spin injection into TSC $\beta$-PdBi$_2$ thin films at room temperature. Instead of using bulk crystals grown via a melt-growth method, we use the molecular beam epitaxy (MBE) method for the $\beta$-PdBi$_2$ films growth and fabrication of heterostructure Fe/$\beta$-PdBi$_2$. Compared with the melt-growth method, MBE method allows us to grow $\beta$-PdBi$_2$ films from a single triple layer up to the dozens of triple layers. The ability to control thickness of $\beta$-PdBi$_2$ films at atomic layers level via MBE opens the promising possibilities to explore fascinating spin transport properties of this quantum material. From the spin pumping measurements we derive a spin diffusion length of 1.76 nm in $\beta$-PdBi$_2$ and an interfacial spin-mixing conductance $4.07 \times 10^{19}\ m^{-2}$. Importantly, the room temperature value of the spin Hall angle 0.037 in $\beta$-PdBi$_2$ was determined, highlighting a good spin to charge conversion efficiency. Our results provide insight into the spin transport in $\beta$-PdBi$_2$, opening the door for its use in superconducting spintronic devices.

## II. EXPERIMENTAL DETAILS

Fe (12 nm) thin films were deposited on MgO (001) substrates by MBE with a base pressure of $3\times10^{-10}$ mbar. Prior to deposition substrates were heated in situ at 600 °C for 2 h to remove surface contaminations. $\beta$-PdBi$_2$ films were then grown on top of the Fe layer by co-evaporating high-purity elemental Bi (99.999%) and Pd (99.99%) via conventional Knudsen cells under Bi-rich condition at 300 °C. The thicknesses of $\beta$-PdBi$_2$ layers vary from 1–12 triple layers (TLs, 1 TL=0.65 nm). Meanwhile, a Fe (12 nm)/Cu (2 nm) bilayer was also fabricated as a reference for the evaluation of the Gilbert damping enhancement and spin pumping induced voltage due to the $\beta$-PdBi$_2$ spin sink layer. The crystallographic properties of the films were examined by grazing angle x-ray diffraction (XRD). The thicknesses and surface morphology were accurately determined by x-ray reflectivity (XRR) measurements and atomic force microscopy (AFM). The static magnetic characteristics were investigated by a Magnetic Property Measurement System (MPMS, Quantum Design). Room temperature FMR and spin pumping measurements were performed using commercial coplanar waveguides and microwave signal generators allowing for lock-in measurements, and the corresponding magnetic field directions were fixed in the Fe film plane along the hard axes. Transport measurements of unpatterned samples with bare $\beta$-PdBi$_2$ and a Fe/$\beta$-PdBi$_2$ bilayer were performed using the standard four-probe method in a Physical Properties Measurement System (PPMS; Quantum Design).

## III. RESULTS AND DISCUSSION

Tetragonal $\beta$-PdBi$_2$ crystallizes into the CuZr$_2$-type structure with a space group of I4/mmm, with the lattice parameters $a = b = 3.36$ Å and $c = 12.98$ Å, as shown in Fig. 1(a). The atomic structure can be visualized as a superposition of triple layers (TLs) with a Bi–Pd–Bi sequence at the center of the unit cell along the $c$-axis. Within each triple-layer, every Pd atom occupies the center of a cube formed by 8 Bi atoms. The XRD patterns of $\beta$-PdBi$_2$ crystal and Fe/$\beta$-PdBi$_2$ bilayer are presented in Fig. 1(b). Only (00l) diffraction peaks are visible in the patterns, confirming high quality $c$-axis crystallographic orientation of the as-grown $\beta$-PdBi$_2$ samples. Film thickness and roughness for single component films in Fe/$\beta$-PdBi$_2$ were established by XRR, as shown in Fig. 1(c). Figures 1(d) and 1(e) present an AFM image of the bare Fe (12 nm) film grown on the MgO substrate and an AFM image of the subsequently grown $\beta$-

PdBi$_2$ (3 TLs) film, respectively. The AFM characterization of the Fe film shows that the film is smooth and has a root-mean-square (RMS) surface roughness of 0.13 nm, while the AFM measurement of the $β$-PdBi$_2$ film shows that the film is granular, with an RMS roughness value of about 0.92 nm. These values are also consistent with the results evaluated from XRR, which shows the RMS roughness of 0.15 nm and 0.89 nm for the Fe and $β$-PdBi$_2$, respectively. Fig. 2 (a) describes the temperature dependence of the longitudinal resistance of the samples 5 TLs $β$-PdBi$_2$ and Fe (12 nm)/$β$-PdBi$_2$ (5 TLs) from 2 to 300 K with no applied field. For the bare $β$-PdBi$_2$, one can see a clear transition to the superconducting state at critical temperature $T_c$ = 4.0 K, which is slightly lower than that for the bulk $β$-PdBi$_2$ with $T_c$ = 5.4 K [11]. However, when compared with thin films grown on a Si substrate with a $T_c$ = 3.1 K [25], the $β$-PdBi$_2$ thin films in our study show better superconductivity due to the higher quality of samples. By adding an Fe layer, the $T_c$ in the $β$-PdBi$_2$ sample is strongly suppressed by the inverse proximity effect [26]. As the superconducting proximity effect leads to leakage of superconducting correlations into the neighboring FM layers. However, for the Fe/$β$-PdBi$_2$ structure, the superconductivity is not annihilated and just the critical temperature $T_c$ decreases below 4 K. In addition, these films were also characterized by the MPMS magnetometer, which shows that the Fe layers have a saturation magnetization of 1.93 T at room temperature as shown in Fig. 2(b), in good agreement with previously determined values for Fe films [27]. The saturation magnetization determined in this way, as well as the coercivity shown in the inset of Fig. 2(b), indicate a weakly $β$-PdBi$_2$ thickness dependent, further confirming the high quality nature of the films.

To investigate the spin dynamics in $β$-PdBi$_2$, we carried out ferromagnetic resonance (FMR) measurements as illustrated in Fig. 3(a). Fig. 3(b) shows the FMR spectra measured for the single-layer Fe (12 nm) and Fe (12 nm)/$\beta$-PdBi$_2$ (5 TLs) bilayer at a fixed frequency of 9 GHz, from which the resonant field $H_{res}$ and linewidth $\Delta H$ can be obtained. A marked enhancement in FMR spectral linewidth after the $\beta$-PdBi$_2$ growth can be observed, which indicates that $\beta$-PdBi$_2$ introduced additional damping in the Fe layer attributed mainly to its function as a spin sink, as described by the spin pumping effect. The inset in Figure 3(b) shows the frequency dependence of the resonant field for the bare Fe and Fe/$\beta$-PdBi$_2$(5 TLs) films, where the effective magnetization $M_{eff}$, g-factor and magnetic anisotropy field $H_k$ of the Fe layer can be extracted using the Kittel formula (see Supplemental Material [28]). The cubic anisotropy field $H_k$ and g-factor, which are both independent of the $\beta$-PdBi$_2$ thickness, are 60 mT and 2.09, respectively. We note here that the effective magnetization ($\mu_0 M_{eff}$, 2.0 T) extracted from the FMR for Fe/$\beta$-PdBi$_2$ bilayers shows slightly larger values than the saturation magnetization ($\mu_0 M_s$, 1.93 T) measured by MPMS (see Fig. 2(b)). This indicates the presence of a negative uniaxial anisotropy, $H_{int} = M_s - M_{eff}$, with the negative sign signifying that it favors an in-plane magnetization. This enhanced $M_{eff}$ has often been observed in FM/TI heterojunctions (like YIG/(Bi$_x$Sb$_{1-x}$)$_2$Te$_3$) at room temperature due to the presence of Dirac surface states [29]. Recently, studies have revealed the SOC-induced spin textures of $\beta$-PdBi$_2$ and shown that not only the topological surface state, but also all other trivial surfaces including part of the bulk bands, exhibit spin polarization parallel to the surface due to the in-plane inversion symmetry [13,14]. These spin-polarized states may potentially couple to a neighboring FM layer and induce an additional in-plane magnetic anisotropy that enhances $M_{eff}$ in our experiments. In the FM/TI heterostructures, the out-of-plane ferromagnetic order can affect the topological surface state, potentially leading to a gap opening, as a result, this will affect the spin pumping especially when the Fermi level is within the band gap [30]. However, in our case, the possible change of the surface state due to Fe layer plays a minor role in the present study since the Fe layer is usually in-plane magnetized [31,32]. In other words, once the temperature reaches below the $T_c$, the topological superconducting phase of the $\beta$-PdBi$_2$ still appear.

The effective Gilbert damping parameter $\alpha_{\text{eff}}$ is extracted by a linear fit of the FMR linewidth $\Delta H$ versus frequency $f$:

$$\mu_0 \Delta H = \mu_0 \Delta H_{\text{inh}} + \frac{4\pi f \alpha_{\text{eff}}}{\gamma} \tag{1}$$

where $\gamma$ is the gyromagnetic ratio and $\Delta H_{\text{inh}}$ is the inhomogeneous broadening. As shown in Fig. 3(c), the linear dependence of $\Delta H$ on $f$ indicates that the damping is reliable and free from extrinsic contributions like two-magnon scattering [33]. The resulting value of effective damping in the Fe/$\beta$-PdBi$_2$ (5TLs) bilayer is found to be 0.0077, which is much higher than that of the bare Fe film (0.0043) indicating the occurrence of spin pumping. Additionally, the measured effective damping as a function of $\beta$-PdBi$_2$ film thickness is plotted in Fig. 3(d). The samples with a $\beta$-PdBi$_2$ thickness larger than 5 TLs exhibit almost the same damping constants, which indicates that the spin pumping induced damping enhancement has saturated. Note that the trend in Fig. 3(d) is same as that observed in normal metal (NM)/ferromagnetic metal (FM) structures, in which the effective spin-mixing conductance $g_{\text{eff}}^{\uparrow\downarrow}$ increases with increasing NM thickness as a result of vanishing spin backflow in thicker NM [34]. Based on the theory of spin-pumping, the flow of angular momentum across the FM/NM bilayer interface is determined by the spin-mixing conductance $g^{\uparrow\downarrow}$ and spin diffusion length $\lambda_{\text{sd}}$, which can be obtained by fitting the $t_{\beta-\text{PdBi}_2}$-dependent Gilbert damping with the relation [34]:

$$\alpha_{\text{eff}} = \alpha_{\text{Fe}} + g^{\uparrow\downarrow} \frac{g\mu_B}{M_s t_{\text{Fe}}} (1 - e^{-(2t_{\beta-\text{PdBi}_2})/\lambda_{\text{sd}}}), \tag{2}$$

where the effective Gilbert damping, $\alpha_{\text{eff}}$, includes the spin-pumping contribution from the Fe/$\beta$-PdBi$_2$ bilayer, in addition to the intrinsic value of the bare Fe film, $\alpha_{\text{Fe}}$. Here, $\mu_B$, $t_{Fe}$, and $t_{\beta-\text{PdBi}_2}$ are the Bohr magneton, Fe thickness, and $\beta$-PdBi$_2$ layer thickness, respectively. The best fit to the data in Fig. 3(d) using Eq. (2) yields values for $g^{\uparrow\downarrow}$ and $\lambda_{\text{sd}}$ at room temperature of $4.07\times10^{19}$ m$^{-2}$ and 2.72 TLs (1.76 nm), respectively. The magnitude of the spin mixing conductance is comparable to those recently reported in FM/Pt(Pd) thin films, such as Fe/Pt(Pd) (4.4(1.37)$\times10^{19}$ m$^{-2}$) [35,27], YIG/Pt (9.7$\times10^{18}$ m$^{-2}$) [36], and Py/Pt (2.53$\times10^{19}$ m$^{-2}$) [5], which implies an excellent spin sink behavior of the $\beta$-PdBi$_2$ and an efficient spin pumping across the bilayer interface.

To evaluate the spin-to-charge conversion efficiency, we measured the transverse electric voltage produced under the FMR condition in Fe/$\beta$-PdBi$_2$ bilayers as a result of the ISHE. Figure 4(a) shows the room temperature spectra of the spin-pumping voltage $V_{\text{SP}}(H)$ measured at a frequency of 9 GHz and microwave power of 50 mW for the bilayer Fe (12 nm)/$\beta$-PdBi$_2$ (5 TLs). The voltage signal changes its sign with reversal of the magnetic field polarity, corresponding to a change in the spin polarization direction of the generated spin current, confirming its ISHE origin [37]. We can eliminate possible parasitic contributions from anisotropic magnetoresistance (AMR) and anomalous Hall effect (AHE), as discussed in Supplemental Material [28]. We also performed a control experiment using a Fe/Pd bilayer sample in order to compare the spin-to-charge conversion efficiency with $\beta$-PdBi$_2$, as well to confirm the accuracy of $V_{\text{ISHE}}$ extraction (see Supplemental Material [28]). Figure 4(b) shows the spectra of the spin-pumping voltage measured at different microwave powers at a constant excitation frequency. Additionally, we plot microwave power $P$ dependence of the $V_{\text{ISHE}}$ in the inset of Fig. 4(b). The $V_{\text{ISHE}}$ increases linearly with $P$, further confirming that the extracted $V_{\text{ISHE}}$ is induced by spin pumping from ISHE [38] and demonstrating that $\beta$-

PdBi$_2$ is able to efficiently interconvert spin and charge currents. For a 3D-SOC system, the electric voltage due to the ISHE is characterized by the spin Hall angle $\theta_{SH}$ [39], which can be described as:

$$V_{ISHE} = \theta_{SH} L R \lambda_{sd} \tanh\left(\frac{d_{\beta-PdBi_2}}{2\lambda_{sd}}\right) \times J_S \tag{3}$$

$$J_S = \frac{g_{eff}^{\uparrow\downarrow}\hbar}{8\pi}\left(\frac{\mu_0 h_{rf}\gamma}{\alpha_{Fe/\beta-PdBi_2}}\right)^2 \left(\frac{\mu_0 M_{eff}\gamma + \sqrt{(\mu_0 M_{eff}\gamma)^2 + 4\omega^2}}{(\mu_0 M_{eff}\gamma)^2 + 4\omega^2}\right)\frac{2e}{\hbar} \tag{4}$$

where $\omega(2\pi f)$, $h_{rf}$, $\hbar$, and $e$ are the excitation frequency, microwave rf field (0.086 mT for 50 mW), Planck's constant, and electronic charge, respectively. $R$ and $L$ are the sample resistance and the width of the microwave transmission line (0.2 mm) for the unpatterned samples. We can experimentally quantify the effective spin mixing conductance as $g_{eff}^{\uparrow\downarrow} = \frac{4\pi M_s t_{Fe}}{g\mu_B}(\alpha_{Fe/\beta-PdBi_2} - \alpha_{Fe})$. Using the spin transport parameters discussed above, the free parameters of $\theta_{SH}$ and $\lambda_{sd}$ are extracted by fitting the thickness dependence of the produced charge current as shown in Fig. 4(c) using Eq. (3), where the best fitting was achieved for 0.037 and 2.50 TLs (1.63 nm), respectively. The obtained $\lambda_{sd}$ agrees very well with the one extracted from the thickness dependence of Gilbert damping via FMR. Compared with the spin diffusion length of bare Pd (7 nm) and Bi (20 nm) [5, 40], the $\lambda_{sd}$ of $\beta$-PdBi$_2$ appears to be much shorter, and is comparable to that of Pt [35] and some antiferromagnets [41,42]. The measured value of $\theta_{SH}$ of $\beta$-PdBi$_2$ is almost one order of magnitude larger than that of Pd alone ($\theta_{SH} = 0.0059$, see Supplemental Material [28]). The other component, Bi, a small gap semimetal with large SOC and one of the most commonly used elements in forming topological materials, has been reported to have an extremely low spin-to-charge conversion efficiency compared to the 5d transition metals shown in the Fig. 4(d) [40]. Fig. 4(d) also shows a useful comparison of spin Hall angles between $\beta$-PdBi$_2$ and other materials. Compared with the superconductors (Nb [23] and NbN [24]) in the normal state and TI Bi$_2$Te$_3$ [43], it can be seen that $\beta$-PdBi$_2$ possesses an even larger conversion efficiency. The obtained $\theta_{SH}$ of $\beta$-PdBi$_2$ is also comparable to the values reported for a commonly used TI (Bi$_2$Se$_3$ [44,45]) and the best spin Hall metals (like Ta [3] and Pt [5]), which indicates that $\beta$-PdBi$_2$ is a good spin current detector. Meanwhile, for comparison purposes, in $\beta$-PdBi$_2$ one can convert spin Hall effect into an equivalent efficiency $\lambda_{IREE}$ through $\lambda_{IREE} = \theta_{SH}\lambda_{sd} \approx 65$ pm, which is comparable to that of superconductor Bi/Ni nanowires [46].

According to the band texture of $\beta$-PdBi$_2$, many interesting topological states exist in this system. Besides the presence of Dirac cone electronic dispersion states at higher-binding energy, topologically protected Rashba surface states were also experimentally identified via ARPES [12]. The observed spin-polarized surfaces thus strongly remind us of the helical edge states in three-dimensional (3D) strong TIs and at Rashba interfaces. As we know, the resonant precession of the magnetization in the FM layer transfers spin angular momentum into the surface states, also accompanied by enhanced damping or a dc voltage via the inverse IREE Rashba-Edelstein effect (IREE). To be noted, the mechanisms of the spin-to-charge conversion related to the ISHE and the IREE are essentially different. For the IREE induced via interfacial Rashba SOC, a charge current converted from a pumped spin current does not depend on the thickness of the spin sink layer. By contrast, in the bulk-SOC induced ISHE, the length scale of the conversion is determined by the spin diffusion length. Intriguingly, the spin pumping induced charge-current presented in Fig. 4(c) shows hyperbolic tangent behavior as a function of the film thickness. This dependence is quite different from that in FM/TI heterostructures as expected from the IREE [47,48] but tracks the behavior observed in heavy metals in accordance with the ISHE. This suggests that a significant amount of charge current is generated from the bulk of the alloy meaning that SOC is dominant in interconversion at room temperature, which has also been seen in other topological materials like Bi-Sb alloys [49] and WTe$_2$ [50]. Here we note that spin-polarized topological surface states are derived as a consequence of SOC in $\beta$-PdBi$_2$, and band inversion is induced by Bi6p–Pd4d mixing. Like some metal alloys with a sizable spin-orbit effect [42,51,52], the larger spin-to-charge conversion in $\beta$-PdBi$_2$ may be mainly due to the transfer of SOC through orbital hybridization.

Even so, the coupling of the two surface states with opposite spin polarizations may decrease the interfacial spin momentum "locking" efficiency at low TL numbers, which results in the decrease of charge current. But we think this possibility is low. It is important to note that an enhancement of damping in the large Bi$_2$Se$_3$ thickness regime is not observed considering the coupling of two surface states [45]. In our case the dependence of damping on the film thickness shown in Fig. 3(d) shows hyperbolic tangent behavior, which is typically observed in the heavy metals due to the decrease in backflow spin current caused by the spin diffusion in the bulk. This also confirms the bulk SHE dominated spin-charge conversion mechanism. Meanwhile, for the non-trivial surface states of $\beta$-PdBi$_2$, a Dirac cone and Rashba-like crossing point appears approximately 2.4 eV below and 2.26 eV above the Fermi level, respectively. Though the coupling of two surface states exists, under a multichannel scattering condition, the surface states far away from the Fermi level do not significantly contribute to the interfacial spin accumulation that bridges the interconversion of spin current and charge current, in accordance with recent numerical calculations [53]. Also, recent studies show that part of the spin-polarized surface states near the Fermi energy are found to be nonhelical in nature as a result of interlayer and intralayer hopping in the spatial distribution of the local spin states [14]. These nonhelical textures may play a minor role in the spin-to-charge conversion via IREE. In TI/FM heterostructures, tuning of the Fermi level position via doping in TI thin films is critical in maximizing the spin-charge conversion efficiency using the spin–momentum locking mechanism [54]. The similar effect may appear in the doped $\beta$-PdBi$_2$. Recent works have revealed evidence for a possible spin excitation associated with Fermi surface reconstruction for the doped $\beta$-PdBi$_2$ [55,56]. Once the Fermi level shifts towards the Dirac cone or Rashba-like crossing point, the spin-charge conversion efficiency may increase due to the enhanced interfacial spin accumulation in FM/doped $\beta$-PdBi$_2$ systems.

While a more quantitative analysis of its non-trivial surface states will be needed to separate out these effects, detailed characterizations of the spin Hall phenomena in the FM/$\beta$-PdBi$_2$ at the superconducting state (below 4 K) would be pursued, at which Majorana modes in $\beta$-PdBi$_2$ may be captured in the presence of the non-trivial superconducting excitations. This may provide an opportunity to investigate the relationship between superconductivity, topology, and even the Majorana fermion. Like some FM/SC structures, quasiparticle-mediated spin-to-charge interconversion would drops rapidly below $T_c$ due to the strong decay of the charge-imbalance relaxation length across $T_c$ [24]. However, as mentioned above, observation of half-quantum flux verifies unconventional superconductivity of $\beta$-PdBi$_2$, consistent with a spin-triplet pairing symmetry. Accordingly, the spin-to-charge conversion may be enhanced within the topological superconducting situation since spin current can be mediated by spin triplet pairs (spin-triplet supercurrent) in $\beta$-PdBi$_2$ beyond conventional electron-mediated spin current [18], which is also interesting for future spintronics applications. This scenario should be further clarified via experiments. But the measurement would be challenging due to the thermal fluctuation brought by the microwave excitation.

## IV. CONCLUSIONS

We report the observation of a large spin-to-charge conversion efficiency in the Fe/$\beta$-PdBi$_2$ heterojunction at room temperature, with a spin Hall angle and spin diffusion length of $\theta_{SH} = 0.037$ and $\lambda_{sd} = 1.76\ nm$, respectively. We find that the emergent topological superconductor $\beta$-PdBi$_2$ may be employed as a promising material candidate for energy-efficient spin-to-charge transduction at room temperature because of the unprecedently large spin-orbit coupling stemming from its unique electronic band structure. Further studies are needed to fully elucidate the spin conversion physics in this fascinating material system, including a probe of low-temperature properties. Our results suggest that spin currents can act as a sensitive probe to detect a variety of non-trivial surface states, shedding light into understanding the spin-momentum locking properties of newly emergent quantum materials.

**ACKNOWLEDGMENTS**

This work is supported by the National Key Research Program of China (Grant Nos. 2016YFA0300701, and 2017YFB0702702), the National Natural Sciences Foundation of China (Grant Nos. 1187411, 51427801, and 51671212) and the Key Research Program of Frontier Sciences, CAS (Grant Nos. QYZDJ-SSW-JSC023, KJZD-SW-M01, and ZDYZ2012-2). S.Y. E.V. and D.S. are grateful for support from the startup provided by North Carolina State University and NC State University-Nagoya Research Collaboration Grant. E.V., and D.S., also acknowledge the partial support from US National Science Foundation, ECCS-1933297 for data analysis.

**Figures and captions**

**Figure 1**

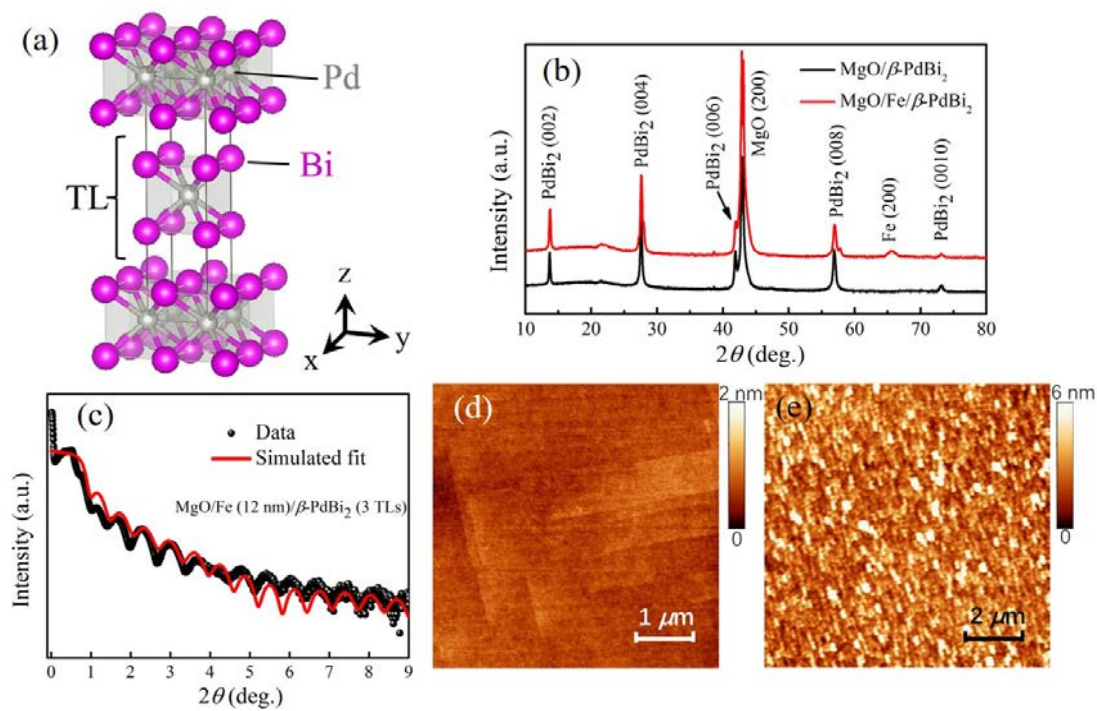

**Fig. 1 (color online)** (a) Schematic illustration of the crystal structure of $\beta$-PdBi$_2$. (b) X-ray diffraction spectra obtained for $\beta$-PdBi$_2$ (5 TLs) and Fe (12 nm)/$\beta$-PdBi$_2$ (5 TLs) on an MgO substrate grown by the MBE method, respectively. Both samples show the expected structure with a preferred orientation along the c axis. (c) Obtained XRR (black) and fitted curve (red) for the Fe (12 nm)/$\beta$-PdBi$_2$ (3 TLs) sample, by which the thickness of Fe layer and the top $\beta$-PdBi$_2$ film are determined. (d) and (e) show surface topography AFM images measured for the bare Fe (12 nm) film and after the deposition of the top $\beta$-PdBi$_2$ layer, respectively.

**Figure 2**

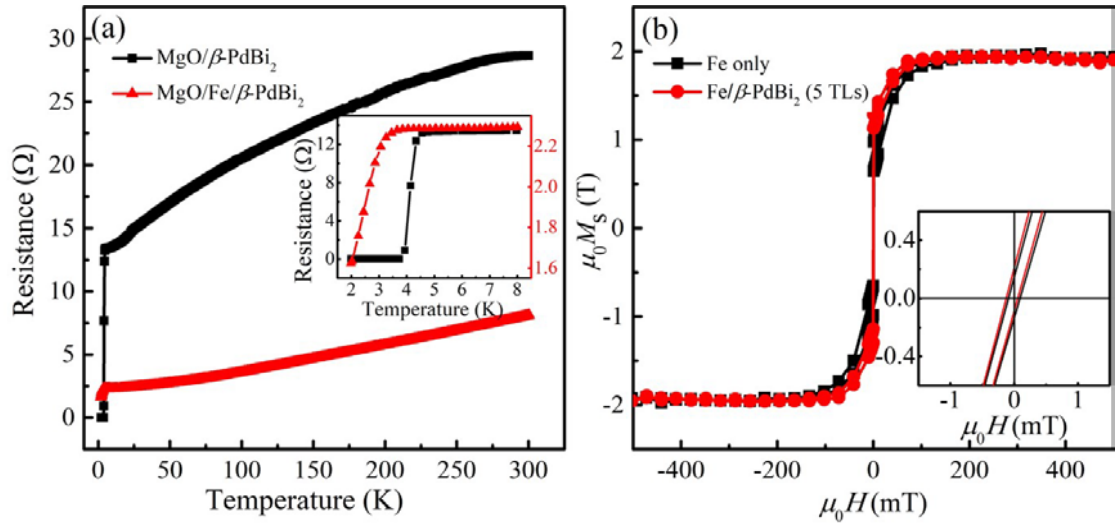

**Fig. 2 (color online)** (a) Temperature dependence of the four-probe resistance in the bare $β$-PdBi$_2$ (5 TLs) film and Fe (12 nm)/$β$-PdBi$_2$ (5TLs) bilayer under zero magnetic field. The inset shows an abrupt change of resistance below $T$=4 K in both samples, indicating the superconducting transition temperatures as reported [11]. (b) Obtained magnetic hysteresis loops of the bare Fe (12 nm) and Fe (12 nm)/$β$-PdBi$_2$ (5TLs) bilayer. The inset is zoomed hysteresis for coercivity. All the measurements were taken at room temperature.

**Figure 3**

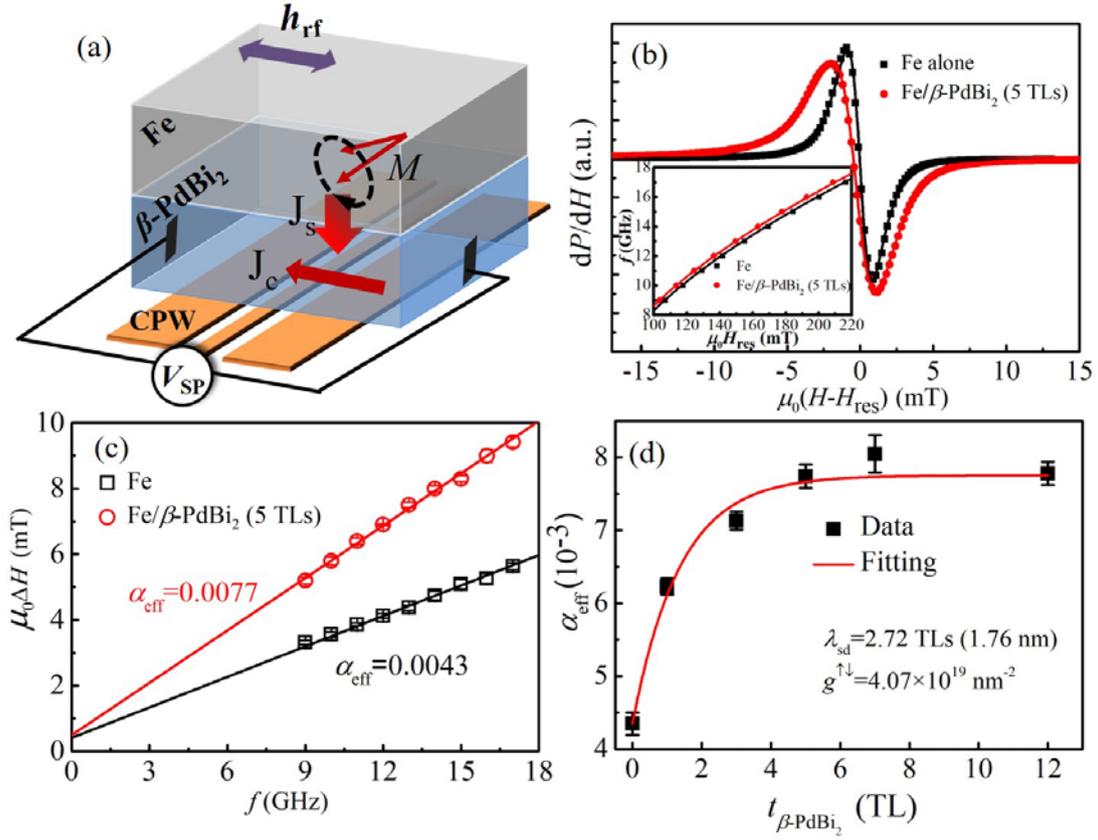

**Fig. 3 (color online)** (a) Schematic illustration of the FMR-driven spin pumping process in the Fe/$\beta$-PdBi$_2$ bilayer. (b) Room temperature FMR spectra of Fe (12 nm)/$\beta$-PdBi$_2$ (5 TLs) and Fe (12 nm) recorded at a microwave frequency $f$=9 GHz. The magnetic field is applied along the hard axes of the film. Solid lines are fitted curves. The inset depicts the plots of microwave frequency $f$ plotted as a function of the resonant field, $H_{res}$ in the bare Fe (12 nm) film, and Fe (12 nm)/$\beta$-PdBi$_2$ (5 TLs) bilayer, respectively, where the solid curve is a fit to the Kittel formula. (c) Frequency dependence of FMR linewidth for the Fe/$\beta$-PdBi$_2$ (5 TLs) bilayers and bare Fe sample at room temperature. From the linear fit using Eq. (1), the damping factor, $\alpha_{eff}$, in each sample can be obtained. (d) $\alpha_{eff}$ plotted as a function of $\beta$-PdBi$_2$ thickness. The red solid curve shows a fitting using Eq. (2), from which the spin diffusion length and the spin mixing conductance are obtained.

**Figure 4**

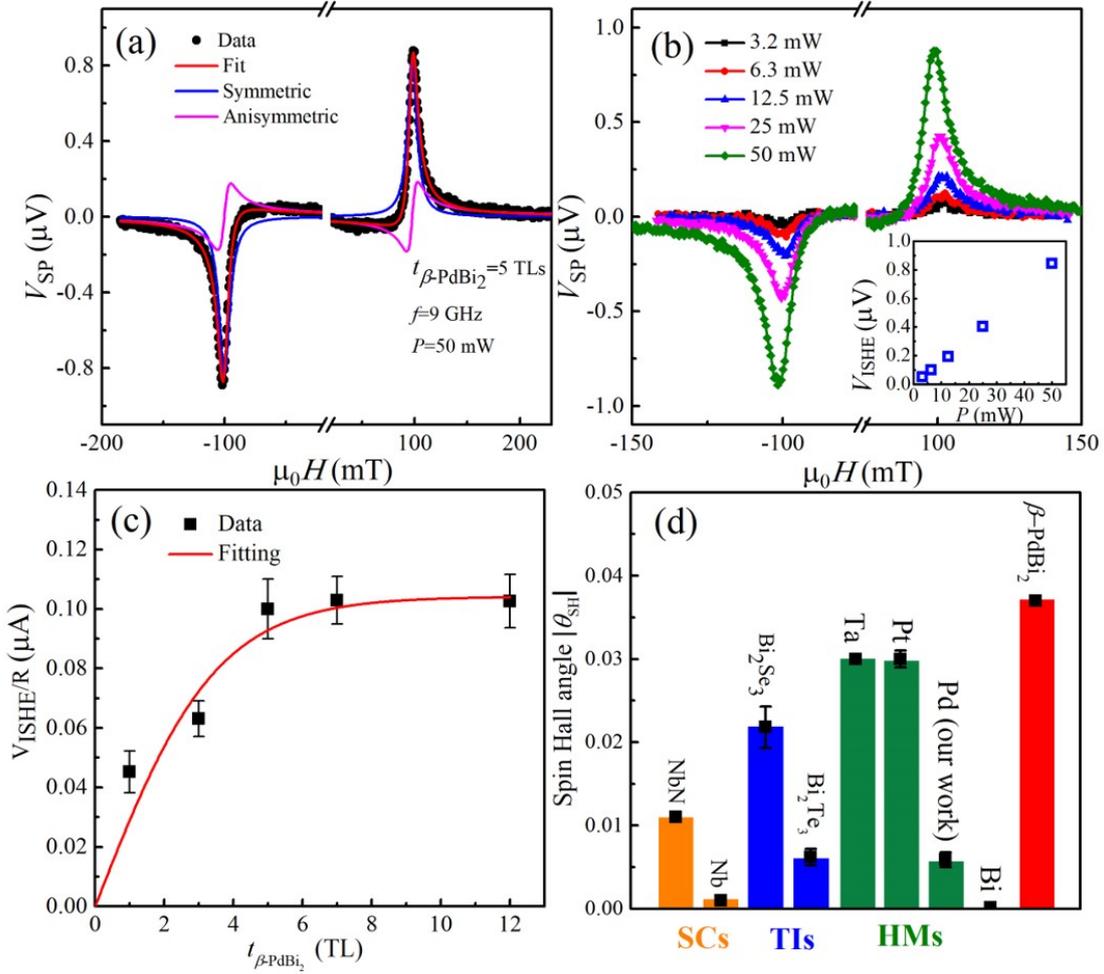

**Fig. 4 (color online)** (a) Room-temperature ISHE voltage measured as a function of the applied magnetic field in the Fe/$\beta$-PdBi$_2$ (5 TLs) sample at a microwave frequency of 9 GHz. The microwave power is fixed at 50 mW. The fitted curves show the separated symmetric (i.e., ISHE) and antisymmetric (i.e., AHE) components, respectively. (b) ISHE voltage measurements under different microwave power ($P$) for the Fe/$\beta$-PdBi$_2$ (5 TLs) sample at $f$=9 GHz. The inset shows the obtained ISHE values as a function of microwave power. (c) Spin pumping induced charge current as a function of $\beta$-PdBi$_2$ thickness. (d) Calculated spin Hall angle ($|\theta_{SH}|$) in the $\beta$-PdBi$_2$ film, compared with the reported $\theta_{SH}$ in the Nb-based superconductors [23,24], topological insulators [43-45], and heavy metals [3,5,40]. Noted that for the uniform comparison, here we just put the values obtained from the spin pumping technic that is same as our measurement.